# First insight into the regime transition dynamics of mode-locked fibre lasers

Guoqing Pu[1], Lilin Yi[1], Li Zhang[1], Yuanhua Feng[2], Zhaohui Li[3], Weisheng Hu[1]

[1]State Key Lab of Advanced Communication Systems and Networks, Shanghai Jiao Tong University, Shanghai, 200240, China. [2]Department of Electronic Engineering, College of Information Science and Technology, Jinan University, Guangzhou 510632, China. [3]School of Electronics and Information Technology, Sun Yat-sen University, Guangzhou 510006, China. (email: lilinyi@sjtu.edu.cn; lzhh88@mail.sysu.edu.cn)

Mode-locked lasers exhibit complex nonlinear dynamics. Precise observation of these dynamics will aid in understanding of the underlying physics and provide new insights for laser design and applications. The starting dynamics, from initial noise fluctuations to the mode-locking regime, have previously been observed directly by time-stretched transform-based real-time spectroscopy. However, the regime transition dynamics, which are essential processes in mode-locked lasers, have not yet been resolved because regime transition process tracking is very challenging. Here we demonstrate the first insight into the regime transition dynamics enabled by our design of a real-time programmable mode-locked fibre laser, in which different operating regimes can be achieved and switched automatically. The regime transition dynamics among initial noise fluctuations, Q-switching, fundamental mode-locking and harmonic mode-locking regimes have been observed and thoroughly analysed by both temporal and spectral means. These findings will enrich our understanding of the complex dynamics inside mode-locked lasers.

The internal dynamics of mode-locked fibre lasers (MLFLs) are attractive for various applications of MLFLs[1] in fields including signal processing[2,3], optical frequency measurement[4,5], ranging metrology[6,7], high-resolution atomic clocks[8,9], and even astronomy[10,11]. Recently, several researchers have investigated these dynamics using the time-stretch dispersive Fourier transform (TS-DFT)[12]. Through a dispersive medium, the TS-DFT builds up a mapping from the spectrum to the temporal domain pulse and a real-time optical spectrum analysis can be performed by combining the TS-DFT with a real-time oscilloscope[13]. The TS-DFT provides endless possibilities for comprehension of ultrafast physical phenomena, including optical rogue waves[14-16], soliton explosions[17,18], mode-locking build-up processes[19-21] and sophisticated soliton dynamics[22-25]. Herink et al. successively resolved the mode-locking build-up process[19] and soliton molecule formation process[22] of a Kerr-lens Ti: sapphire laser. The mode-locking build-up processes in an anomalous dispersion fibre laser[20] and the build-up dynamics of a dissipative soliton[21] have also been observed. In addition, the complete build-up dynamics of soliton molecules[24] and soliton dynamics related to Q-switched instabilities[25] in MLFLs have also been investigated. However, all these observations are without exception based on the pulse build-up processes that are induced by increasing the pump power. There has been no possibility to date of observation of the pulse transition dynamics from one mode-locking regime to another; these transitions are essential physical processes in MLFLs but tracking of the regime transition process is very challenging. Because regime transitions are very difficult to realise using pump power tuning alone. Therefore, studies of the nonlinear dynamics of MLFLs are not yet complete.

Nonlinear polarization evolution (NPE) is the main method that is used to achieve mode locking. In addition to the fundamental mode-locking (FML) regime, NPE-based MLFLs can also produce harmonic mode-locking (HML) regime with high repetition rates[26] and Q-switching (QS) regime with high pulse energies[27,28] by polarization tuning alone. Therefore, NPE-based MLFLs represent the perfect platform for investigation of the regime transition dynamics. We have previously designed the first real-time programmable MLFL using intelligent polarization tuning that can be automatically locked on and switched among the QS, QML, FML and HML regimes[29]. In this paper, by virtue of use of this real-time programmable MLFL with a combination of temporal and spectral observations obtained using the TS-DFT and a real-time oscilloscope, we first reveal the QS and mode-locking build-up processes induced by polarization tuning. Furthermore, we provide the first insight into the regime transition dynamics among the QS, FML and HML regimes. Our observations have demonstrated some interesting findings that have not been reported previously. For example, the regime transition from QS to stable FML experiences dual-wavelength competition. The formation processes for QS pulses are quite similar, regardless of whether they are built up from the initial noise fluctuations or transited from the FML regime, in which dual-envelope competition always exists and only the dominant pulse can eventually be oscillated. The transition from FML to the HML regime will require tens of thousands of roundtrip times to achieve stable spectral operation, but the reverse transition only requires tens of roundtrip times to achieve stable operation, corresponding to a regime switching time as fast as tens of microseconds. We believe that these interesting findings will provide researchers with a better understanding of the internal dynamics of MLFLs and thus aid in the design and application of these lasers.

## Results

**Experimental setup.** Figure 1 illustrates the complete experimental setup, where the left side shows the real-time programmable MLFL that was enabled by our proposed intelligent algorithm[29]; the fundamental repetition rate of the laser is 4.325 MHz, which corresponds to a roundtrip time of approximately 0.23 μs. An electronic polarization controller (EPC) with an average response time of approximately 5 μs, which is a core component of the laser, is controlled through four DC voltage channels. Thus, by recording several sets of voltages that lead to the different operating regimes achieved using the intelligent algorithm, i.e., the previously experienced values, the laser can switch directly among multiple operating regimes. A chirped fibre Bragg grating (FBG) with a dispersion value of −1651 ps/nm in the C band is used for mapping from the spectrum to the temporal domain pulse (see Methods for details). Undispersed waveforms are also recorded using the real-time recording branch. Notably, to ensure successful acquisition of each transition, a field-programmable gate array (FPGA) with the intelligent algorithm embedded in it is programmed to transmit a trigger signal to the oscilloscope immediately after the voltages are loaded.

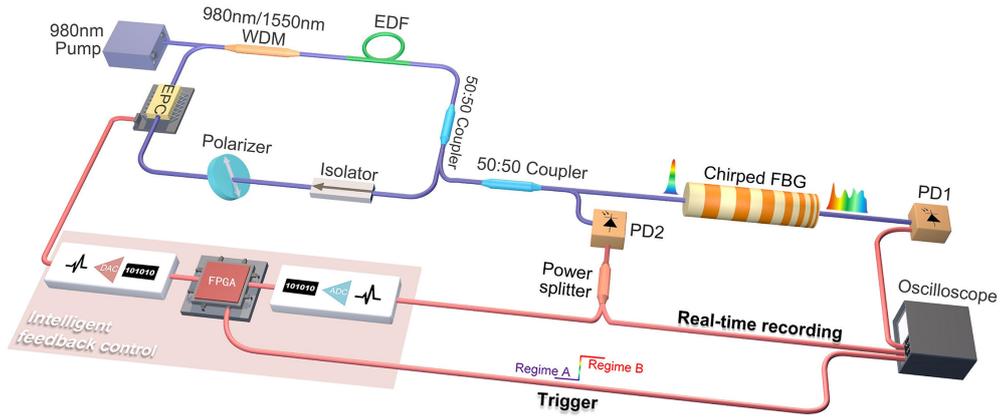

**Fig. 1 Experimental setup.** The left part shows the nonlinear polarization evolution (NPE)-based real-time programmable MLFL. A 980 nm laser pumps an erbium-doped fibre (EDF) as the gain medium. The isolator guarantees unidirectional running and the polarizer is the key component for NPE-based mode-locking. The electronic polarization controller (EPC) is used for polarization tuning. An optical coupler retains half the power inside the cavity while the other half is for output. This output is sequentially divided in two by another optical coupler. The upper branch is used for the TS-DFT through a chirped fibre Bragg grating (FBG). Photodetector 2 (PD2) receives the undispersed signal from the lower branch. After PD2, the left branch is sent to the intelligent feedback panel, consisting of an analogue-to-digital converter (ADC), a field-programmable gate array (FPGA), and a digital-to-analogue converter (DAC) for regime control, while the right branch is sent directly to the oscilloscope for real-time recording.

**Transition dynamics to FML regime from continuous-wave (cw) state and QS regime.**

Figure 2a illustrates the switch from cw state (the initial noise fluctuations) to the FML regime based on loading of the previously experienced voltage value that corresponds to the FML regime into the EPC. From the TS-DFT results (upper part, Fig. 2a), some picosecond fluctuations oscillate rapidly and one of these oscillations eventually leads to mode locking. The obvious boundary in the TS-DFT result and the peak shown in the real-time record (lower part, Fig. 2a) around the 2300th roundtrip arise from the abrupt polarization change produced by the EPC, which is the sign that the experienced values have been loaded. The real-time record shows that the complete transition phase takes approximately 10000 roundtrips, from the 2300th to the 12300th roundtrip, which represents approximately 2.3 ms in our cavity; however, the time cost is rather random and ranges from several hundred microseconds to several milliseconds, even for the identical FML regime during multiple measurements. The initial phase of the transition is shown in Fig. 2b. In the EPC loading phase, there are four larger pulses that correspond to the four channels of the drive voltages loaded on the EPC in series. The intervals between successive pulses are several microseconds long and correspond to the response time of the EPC. After the previously experienced values have been loaded, the laser evolves into a small-amplitude Q-switched mode-locking (QML) -type oscillation, which is the temporal manifestation of superposition of multiple picosecond oscillations[19], and maintains this type of oscillation until it reaches the FML regime. Figure 2c shows the dynamic when approaching the FML regime, where some spikes stand out in the small-amplitude QML-type oscillation. Apparently, these spikes trigger the onset of mode-locking.

Figure 2d illustrates the transition dynamics from QS to the FML regime. The TS-DFT results manifest in an intriguing dynamic after the EPC loading process, where there are initially two dominant wavelengths at 1531 nm and 1542 nm. Before the evolution into the FML regime, the power at the longer wavelength is obviously stronger than that at the shorter wavelength. However, the relationship flips immediately after entering the FML regime. In a sense, this phenomenon demonstrates competition between the dual wavelengths, where the shorter wavelength wins after completion of the 6000th roundtrip, and the spectrum eventually stabilizes. To enhance our understanding of the dual-wavelength competition, we compared the undispersed and dispersed pulses during competition, as shown in Fig. 2e. Interestingly, the dual-wavelength competition situation corresponds to the dual-pulse phenomenon that occurs in the undispersed time domain, and the time interval between two closely-spaced pulses is only approximately one-fifth of a roundtrip. Obviously, the first peak outweighs the second peak in each dispersed pulse in the dual-pulse phenomenon. Because the chirped FBG applied in the TS-DFT has normal dispersion in the C-band, the higher peak corresponds to the longer wavelength, while the other peak corresponds to the shorter wavelength. This conclusion is identical to that drawn from the TS-DFT results. Additionally, according to our multiple measurements, the dual-pulse phenomenon occurs randomly when switching to the FML regime from either the cw state or the QS regime. After the spectrum stabilises, the dual-pulse phenomenon disappears and the laser operates in the stable FML regime, as indicated by Fig. 2f. For comparison, the transition time from the QS regime to the FML regime is much shorter than that from the cw state to the FML regime.

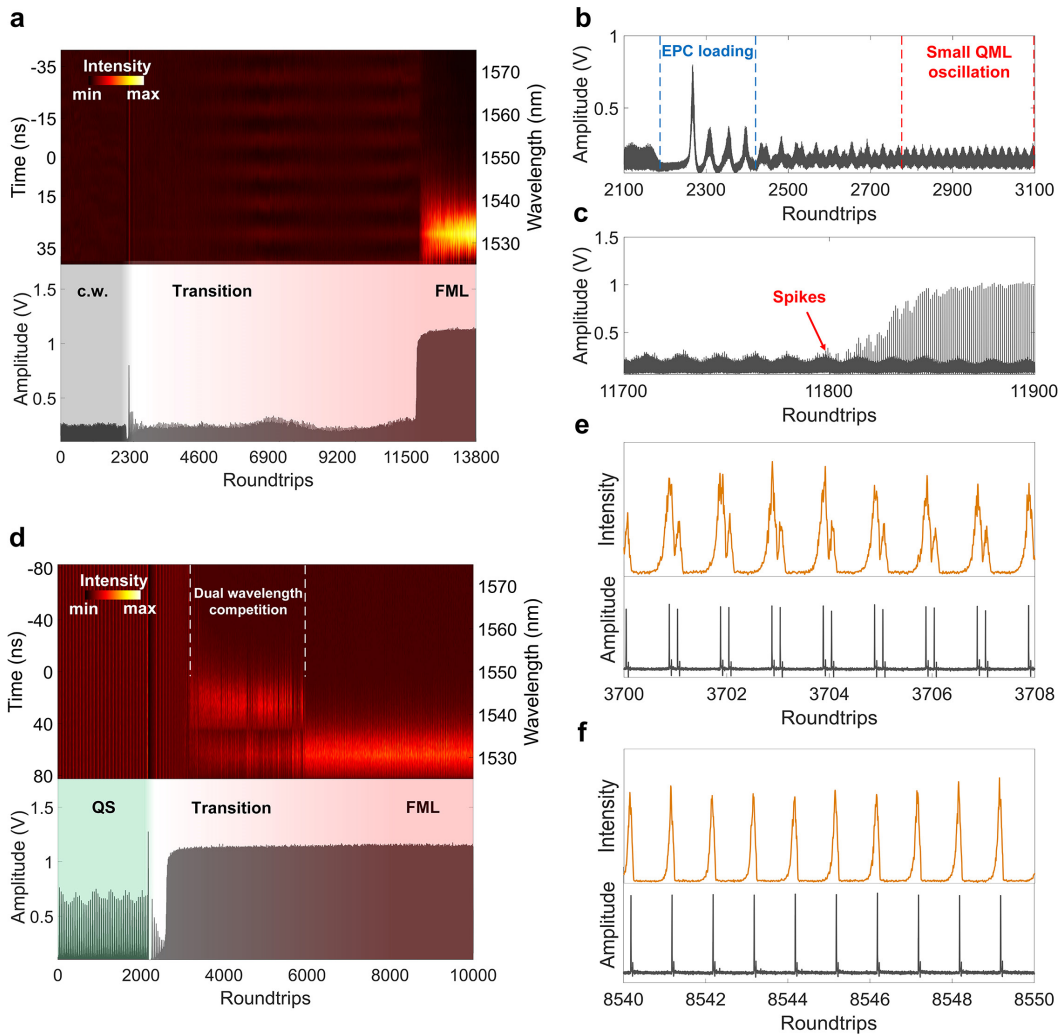

**Fig. 2 Records of switching processes to the FML regime from the cw and QS regimes. a–c,** Records of switching from the cw state to the FML regime. **a,** TS-DFT result (upper) and real-time waveform record (lower). The laser bursts into the FML regime after a lengthy period of oscillation. **b,** The initial part and **c,** the approaching part of the transition phase. **d–f,** Records of switching from the QS regime to the FML regime. **d,** TS-DFT result (upper) and real-time waveform record (lower), dual-wavelength competition rises from approximate the 3000th roundtrip and the shorter wavelength wins out at about the 6000th roundtrip. **e,** Undispersed (lower) and dispersed pulses (upper) of the dual-pulse phenomenon. **f,** Undispersed (lower) and dispersed pulses (upper) under FML regime.

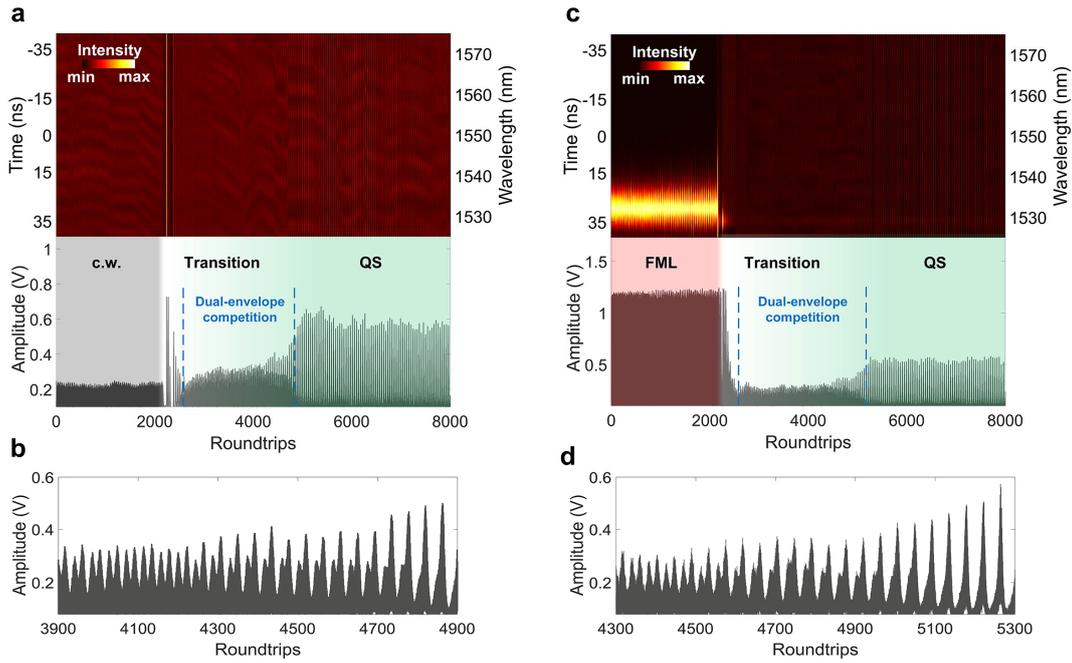

**Fig. 3 Records of switching processes to the QS regime from the cw and FML regimes. a–b,** Records of switching from the cw state to the QS regime. **a,** TS-DFT result (upper) and real-time record (lower). **b,** Approaching part of the transition phase. **c–d,** Records of switching from the FML regime to the QS regime. **c,** TS-DFT result (upper) and real-time record (lower). **d,** Approaching part of the transition phase. Dual-envelope competition is observed in both switching processes to the QS regime.

**Transition dynamics to QS regime from cw state and FML regime.** The transitions to the QS regime from both the cw and FML regimes are recorded in Fig. 3a and c, respectively. The TS-DFT is not actually effective on QS pulses. First, the QS spectrum width is generally narrower than that of the mode-locked pulses, and thus the pulse broadening phenomenon is less obvious. Second, the QS pulse width is much larger when compared with the amount of broadening produced by dispersion. Therefore, the transition analyses of switching to the QS regime are substantially reliant on time-domain records. Analogous to the switching process from the cw state to the FML regime, the laser enters the small-amplitude QML oscillation from both switches. Intriguingly, the period of the envelope of the small-amplitude QML oscillation is almost half of the period of the QS regime. The dynamics when approaching the QS regime from the cw and FML regimes, as illustrated by Fig. 3b and d, respectively, are quite analogous. One of the two envelopes of the small-amplitude QML oscillation is chosen to rise gradually, leading to the desired QS regime. Since this process resembles a sort of competition between these two envelopes, we term it as dual-envelope competition. This peculiar dynamic is also observed in massive switches from different operating regimes or from cw to the QS regime. The transition dynamics and time costs from cw and FML to the QS regime are almost the same.

**Transition dynamics between FML regime and higher-order HML regimes.** The transitions between the FML and higher-order HML regimes are also recorded. The transition phase of the real-time record (Fig. 4b) shows that the transition is completed rapidly, within only 300 roundtrips. Notably, the essential spikes are also observed during the transition phase, thus proving that the mode-locking originates from the spikes that occur inside the QML oscillations. Nevertheless, the spectrum width gradually narrows and ultimately settles down at approximately the 7000th roundtrip, as indicated by the TS-DFT results (upper in Fig. 4a). The undispersed and dispersed pulses before the spectrum is stabilised are shown in Fig. 4c. While the undispersed pulses show stable temporal characteristics in the second-order HML regime, the dispersed pulses are distorted and contain numerous spikes, corresponding to a broad and unstable optical spectrum. Figure 4d shows the undispersed and dispersed pulses after the spectrum stabilised. When compared with the pulses before the spectrum stabilized, the time-domain pulses show little difference, but the dispersed pulses are far smoother and more stable after the spectrum is stabilised. Combining the TS-DFT results with the real-time recording allows us to conclude that the transition from FML to the higher-order FML regimes will require long time to stabilise. Figure 4e shows that the transition from the third-order HML regime to the FML regime is much more direct. There is barely a transition phase from both the TS-DFT results and the real-time record shown in Fig. 4e. However, the spectrum bandwidth broadens greatly when transiting to the FML regime. Figure 4f shows the transition dynamics in detail, from which it is obvious that the spikes that trigger the FML regime appear during loading of the second channel of the EPC and then evolve into a stable FML regime. The switching time from the higher-order HML to the FML regime is of the order of only tens of microseconds, which is much faster than switching from the FML regime to a higher-order HML regime, and is the fastest switching time among the observed regimes. From a comparison of the transition from FML to HML with that from HML to FML, we can conclude that the evolution from the regime with lower stability (i.e., the HML) to that with higher stability (i.e., the FML) is much smoother than the reverse transition, which agrees well with the physical evolution mechanism.

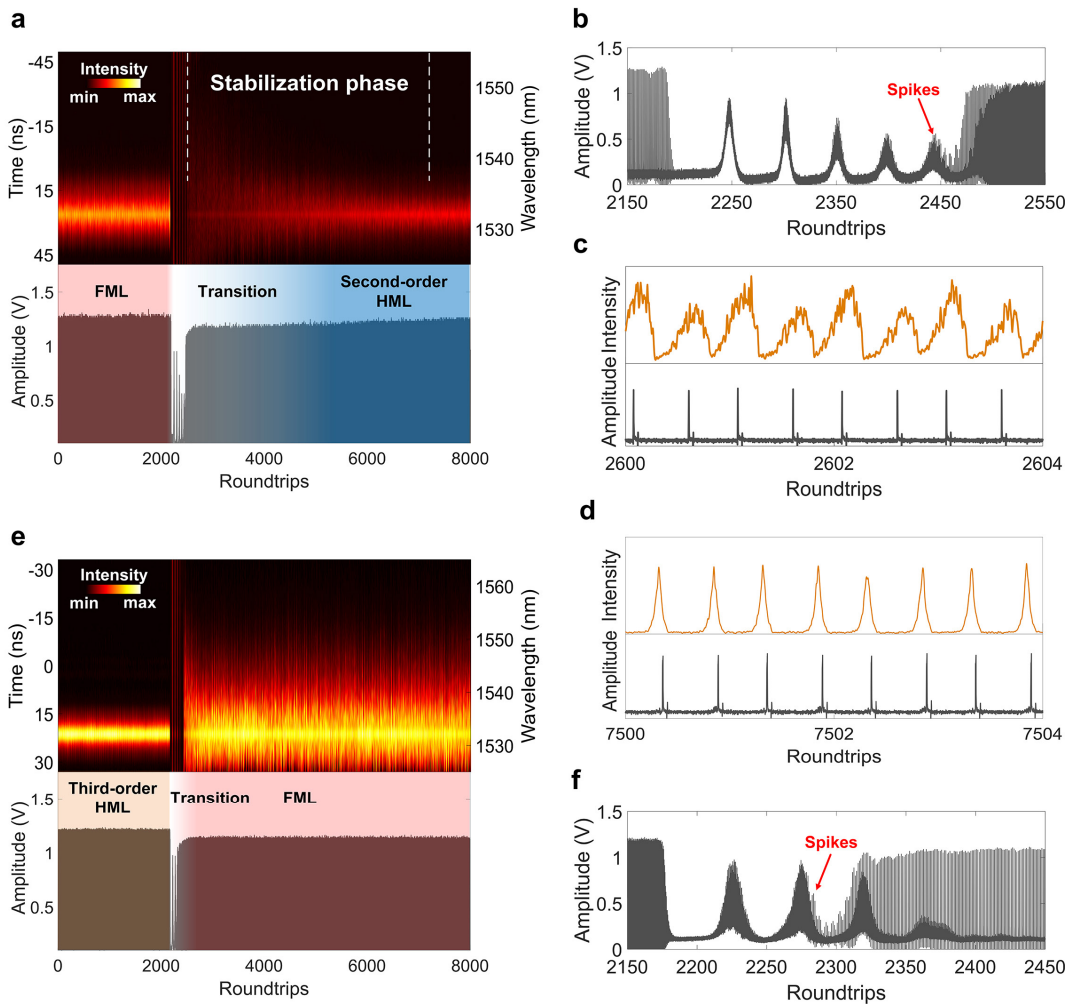

**Fig. 4 Records of switching from the FML regime to the second-order HML regime and switching from the third-order HML regime to the FML regime. a–d,** Records of switching from the FML regime to the second-order HML regime. **a,** TS-DFT result (upper) and real-time record (lower). The spectrum gradually settles down around the 7000th roundtrip. **b,** Transition phase. The undispersed (lower) and dispersed pulses (upper) **c,** before and **d,** after the spectrum stabilises. **e–f,** Records of switching from the third-order HML regime to the FML regime. **e,** TS-DFT result (upper) and real-time record (lower). **f,** Transition phase.

**Discussion**

In conclusion, we first unveiled the dynamics of direct switching among multiple operation regimes by virtue of use of a homemade real-time programmable mode-locked fibre laser[29]. As a result of the insight gained into the regime transition dynamics, we have observed some interesting phenomena that have not been reported previously. The mode-locking behaviour always originated from the spikes that were observed in the small QML oscillations, regardless of whether the transition was from the cw state or QS regimes. The regime transition from QS to the FML regime will occasionally experience dual-wavelength competition. On the other hand, dual-envelope competition always exists and only the dominant pulse can eventually be oscillated for the QS pulse build-up process, no matter whether the transition is from the cw state or FML regimes. The transition time from the HML regime with lower stability to the FML regime with higher stability is much shorter than the reverse process and only requires tens of microseconds, which is the fastest regime transition time reported to date. We believe that these findings can enrich the wider understanding of the complex nonlinear dynamics of mode-locked fibre lasers and will provide new insights for fibre laser design and applications. Utilizing the real-time programmable mode-locked fibre laser will allow us to achieve more mode-locking regimes, including soliton molecule[24,30] regimes. The transition dynamics between these rare regimes will be studied further in future work.

**Methods**

The NPE-based real-time programmable passive MLFL is shown in the left part of Fig. 1. The EPC is driven via four channels with DC voltages of 0–5 V and can respond in microseconds. The bandwidth of PD1 and PD2 is 40 GHz and 10 GHz respectively. The cavity length of the fibre laser is ~47.8 m. Additionally, the intelligent control panel consists of a 400 MSa/s ADC with 8-bit resolution, a Xilinx Zynq FPGA and four 100 MSa/s DACs with 12-bit resolution. The dispersion parameter of the chirped FBG that was used for the TS-DFT is −1651 ps/nm in the C band. Notably, the spectrum width of the input is filtered down to 50 nm (over the range from 1525 nm to 1575 nm) using the chirped FBG. The real-time oscilloscope operates at a sampling rate of 10 GSa/s.

### Acknowledgements
This work is partly supported by National Natural Science Foundation of China (NSFC) (61575122).

### Author contributions
L. Yi conceived the idea. Z. Li discussed the idea with L. Yi and provided the TS-DFT experimental environment. G. Pu designed the real-time programmable mode-locked fibre laser and carried out the experiment with the supervision of both L. Yi and Z. Li. L. Zhang designed and debugged the feedback circuit utilized in the real-time programmable mode-locked laser. Y. Feng assisted to this experiment. All authors discussed the results and contributed to writing the manuscript.

### Competing interests
The authors declare no competing interest.